\documentclass[a4paper,12pt]{article}
\usepackage{epsfig}
\newcommand{\bmat}{\left(\begin{array}}
\newcommand{\emat}{\end{array}\right)}

\def\lsim{\raise0.3ex\hbox{$\;<$\kern-0.75em\raise-1.1ex\hbox{$\sim\;$}}}
\def\gsim{\raise0.3ex\hbox{$\;>$\kern-0.75em\raise-1.1ex\hbox{$\sim\;$}}}

\def\yzero{\smash{\hbox{$y\kern-4pt\raise1pt\hbox{${}^\circ$}$}}}

\def\s2{\frac{1}{\sqrt2}}

\def\beq{\begin{equation}}
\def\eeq{\end{equation}}
\def\beqa{\begin{eqnarray}}
\def\eeqa{\end{eqnarray}}

\def\IF{\relax{\rm I\kern-.18em F}}
\def\II{\relax{\rm I\kern-.18em I}}
\def\IP{\relax{\rm I\kern-.18em P}}
\def\IC{\relax\hbox{\kern.25em$\inbar\kern-.3em{\rm C}$}}
\def\IR{\relax{\rm I\kern-.18em R}}

\def\Dsl{\,\raise.15ex\hbox{/}\mkern-13.5mu D} 
\def\IZ{Z\kern-.4em  Z}
\def\bmat{\left(\begin{array}}
\def\emat{\end{array}\right)}

\def    \part          {\partial}
\def    \be            {\begin{equation}}
\def    \ee            {\end{equation}}
\def    \bea           {\begin{eqnarray}}
\def    \eea           {\end{eqnarray}}

\begin{document}
%
\pagestyle{empty}

\pagestyle{empty}

\rightline{FTUAM 02/31}
\rightline{IFT-UAM/CSIC-02-50}
\rightline{November 2002}

\renewcommand{\thefootnote}{\fnsymbol{footnote}}
\setcounter{footnote}{0}

\vspace{1.5cm}
\begin{center}
\large{\bf Extra Matter at Low Energy\footnote{Talk given at 1st
International Conference on String Phenomenology, Oxford,
6-11 July 2002.}
\\[5mm]}
\vspace{1cm}
\mbox{\sc{
\small{
C. Mu\~noz
}
}
}
\vspace{0.5cm}
\begin{center}
{\small
{\it 
Departamento de F\'{\i}sica
Te\'orica C-XI and
Instituto de F\'{\i}sica Te\'orica  C-XVI,\\
Universidad Aut\'onoma de Madrid,
Cantoblanco, 28049 Madrid, Spain \\
E-mail: carlos.munnoz@uam.es}
}

\end{center}

\vspace{1.5cm}

{\bf Abstract} 
\\[7mm]
\end{center}
\begin{center}
\begin{minipage}[h]{14.0cm}
Assuming that the Standard Model arises
from the $E_8\times E_8$ Heterotic Superstring,
we try to solve the discrepancy between the
unification scale predicted by this theory 
($\approx g_{GUT}\times 5.27\cdot 10^{17}$ GeV) 
and the value deduced from LEP experiments 
($\approx 2\cdot 10^{16}$ GeV). This will allow us
to predict the presence at low energies of 
three generations of supersymmetric Higgses
and vector-like colour triplets. 

\end{minipage}
\end{center}
\vspace{0.5cm}
\begin{center}
\begin{minipage}[h]{14.0cm}

\end{minipage}
\end{center}
\newpage
\setcounter{page}{1}
\pagestyle{plain}
\renewcommand{\thefootnote}{\arabic{footnote}}
\setcounter{footnote}{0}
%
%

\section{Introduction}

Although the Standard Model of particle physics provides
a correct description of the observable world, 
there exist
strong indications that it is just a low-energy effective theory.
There is no answer in the context of the Standard Model to some
fundamental questions.
For example, how can we unify it with
gravity? And then: How can
we protect 
the masses of the scalar particles against quadratic divergences
in perturbation theory (the so-called hierarchy problem)?
Other questions cannot even be
posed:
Why is the Standard Model gauge group
$SU(3)\times SU(2)\times U(1)_Y$?
Why are there three families of particles?
Why is the pattern of quark and lepton masses so weird?

In fact, only Superstring Theory 
has the potential to unify all gauge interactions 
with gravity 
in a consistent 
way being able to answer all the above questions. 
In this sense, it is a crucial step
to build a consistent Superstring Theory in four dimensions
accommodating the observed Standard Model, 
i.e. we need to find the SuperString Standard Model (SSSM).

In the late eighties,
the compactification of the $E_8\times E_8$
Heterotic String 
on six-dimensional orbifolds 
proved to be an
interesting method to carry out this 
task (for a brief historical account see the Introduction in 
ref.~\cite{prediction}
and references therein).
It was shown that the use of two 
Wilson lines 
on the torus defining the
symmetric $Z_3$ orbifold can give rise to four-dimensional
supersymmetric models with gauge group
$SU(3)\times SU(2)\times U(1)^5\times G_{hidden}$
and three generations of chiral particles. 
In addition,
it was also shown 
that the 
Fayet--Iliopoulos D-term, 
which appears because of the presence of 
an anomalous $U(1)$, can give
rise to the breaking of the extra $U(1)$'s.
In this way it was possible to construct \cite{Casas2,Font,Katehou} 
supersymmetric 
models 
with gauge group 
$SU(3)\times SU(2)\times U(1)_Y$,
three generations of particles in the observable sector,
and absence of dangerous 
baryon- and 
lepton-number-violating operators\footnote{Recently
another model has been partially analyzed \cite{Giedt2}.}.
 

Unfortunately, we cannot claim that one of these models is the 
SSSM, since several problems are always present. Let us 
concentrate on two of them. First of all, although 
the initially large number of extra particles,
which are generically present in these constructions, is 
highly reduced through the
Fayet--Iliopoulos mechanism,
since many of them get a high mass 
($\approx 10^{16-17}$ GeV),
some extra $SU(3)$ triplets, $SU(2)$ doublets and 
$SU(3)\times SU(2)$ singlets still remain.
On the other hand,
given the predicted value for the unification scale in
the Heterotic String \cite{Kaplu}, 
$M_{GUT}\approx g_{GUT}\times 5.27\cdot 10^{17}$ GeV,
the values of the gauge couplings deduced from
LEP experiments cannot be obtained \cite{rges}.
Recall that this is only possible 
in the context of the minimal supersymmetric standard
model (MSSM)
for $M_{GUT}\approx 3\times 10^{16}$ GeV.

In any case, it is plausible to think that another orbifold model
could be found with the right properties. In the present talk
we will adopt this point of view, and
we will try to deduce the phenomenological properties 
that such a model must have
in order to solve the two important problems
mentioned above.
In fact, both problems,
extra matter and gauge coupling unification,
are closely related, since
the evolution of the gauge couplings from high to low energy
through the RGEs
depends on the existing matter. 
With our solution 
we will be able to predict
the existence of three generations of supersymmetric Higgses
and vector-like colour triplets
at low energies \cite{prediction}.




\section{Predictions from the unification 
of 
$\alpha_3$ 
with 
$\alpha_2$}


Since we are interested in the analysis of gauge couplings,
we need to first clarify which is the relevant scale for
the running between the supersymmetric scale $M_S$ and the unification
point.
Let us recall that in heterotic compactifications
some scalars singlets $\chi_i$  develop VEVs in order
to cancel the Fayet--Iliopoulos D-term, without breaking the
Standard Model gauge group.
An estimate about their VEVs can be done
with the average 
result $\langle\chi_j \rangle\sim 10^{16-17}$ GeV.
After the breaking, many particles, say $\xi$, acquire a high mass because
of the generation of effective mass terms. These come for example
from operators of the type $\chi_i\xi\xi$.
In this way extra vector-like triplets and doublets and also singlets become 
very heavy. 
We will use the above value as our relevant scale,
the so-called
Fayet--Iliopoulos scale $M_{FI}\approx 10^{16-17}$ GeV.

As discussed in the Introduction, 
we are interested in the unification of the gauge couplings
at
$M_{GUT}\approx g_{GUT}\times 5.27\cdot 10^{17}$ GeV.
Let us try  
to obtain this value by using first the existence of extra matter at the scale
$M_S$. We will see that this is not sufficient and 
the Fayet--Iliopoulos scale 
must be included. 
Let us concentrate for the moment on
$\alpha_3$ and  $\alpha_2$.
Recalling that three generations appear automatically 
for all the matter in $Z_3$ orbifold scenarios with two Wilson lines,
the most natural possibility
is to assume the presence of three light generations of supersymmetric
Higgses.
This implies that we have four extra Higgs doublets, $n_2=4$, 
with respect to the
case of the MSSM. 
Unfortunately, this goes wrong.
Whereas $\alpha_3^{-1}$ remains unchanged, since the number of extra
triplets
$n_3=0$,
the line 
for $\alpha_2^{-1}$ is pushed 
down with respect to the case of the MSSM. 
As a consequence, the two
couplings cross at a very low scale
($\approx 10^{12}$ GeV).
We could try to improve this situation by assuming the presence of
extra triplets in addition to the four extra doublets. 
Then the line for $\alpha_3^{-1}$ is
also pushed down and therefore the crossing might be obtained
for larger scales. However, even for the
minimum number of extra triplets that can be naturally obtained
in our scenario, $3\times \{(3,1)+(\bar 3, 1)\}$, i.e. $n_3=6$,
the ``unification'' scale turns out to be too large ($\approx 10^{21}$
GeV).
One can check that
other possibilities including more extra doublets and/or triplets
do not work \cite{prediction}.
Thus, using extra matter at $M_S$ we are not able to obtain the 
Heterotic String unification scale since
$\alpha_3$ never crosses $\alpha_2$ at
$M_{GUT}\approx g_{GUT}\times 5.27\cdot 10^{17}$ GeV.
Fortunately, this is not the end of the story. 
As we will show now,
the Fayet--Iliopoulos scale $M_{FI}$ is going to
play an important role in the analysis.

In order to determine whether or not the  
Heterotic String unification scale
can be obtained, we need to know the number of doublets
$n_2^{FI}$ and triplets $n_3^{FI}$ 
in our construction with masses of the order of the Fayet--Iliopoulos
scale $M_{FI}$.
It is possible to show
that within
the $Z_3$ orbifold with two Wilson lines, three-generation 
standard-like models
models must fulfil the following relation
for the extra matter:
$2+n_2+n_2^{FI}=n_3+n_3^{FI}+12$.
Then, it is now straightforward to check
that only models with 
$
n_2=4, n_3=6, 
$
%
and therefore 
$n_2^{FI}-n_3^{FI}=12$, may give rise to the Heterotic String
unification scale \cite{prediction} 
(the other possibilities for $n_2$, $n_3$, mentioned
above do not even produce the crossing of $\alpha_3$ and $\alpha_2$). 
This is shown in Fig.~2 for an example with
$n_3^{FI}=0$, and assuming $M_S=500$ GeV. There we are
using $M_{FI}=2\cdot 10^{16}$ GeV as will be discussed below. 

\begin{figure}[t]
\centerline{\epsfxsize=2.5in\epsfbox{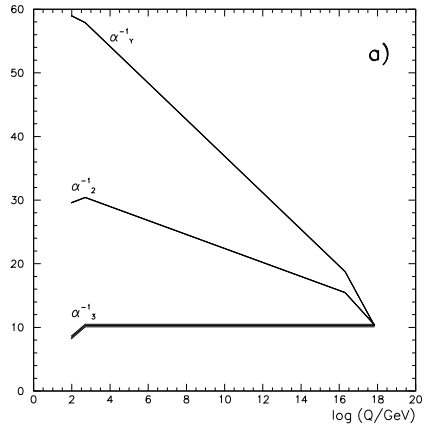}}   
\caption{
Unification of the gauge couplings 
at $M_{GUT}\approx g_{GUT}\times 5.27\cdot 10^{17}$ GeV 
with three light generations of supersymmetric Higgses
and
vector-like colour triplets.
In this example we show one of the four 
possible patterns of heavy matter in eq.~(\ref{extra5}),
in particular that with $n_3^{FI}=0$.
The line corresponding to $\alpha_1$
is just one of the many possible examples.
}
\end{figure}



Note that at low energy we then have (excluding singlets)
\bea
3\times \left\{(3,2)+2(\bar 3,1)+(1,2)\right\}
+ 3\times \left\{(3,1)+(\bar 3,1)+ 2(1,2)\right\}
\ ,
\label{SM2}
\eea
i.e. the matter content of the Supersymmetric Standard Model with
three generations of Higgses and vector-like colour triplets.


Let us remark that 
in these constructions
only the following patterns of matter with masses of the order of $M_{FI}$
are allowed:
%
\bea
a)\ n_3^{FI}=0\ ,\,\ \ n_2^{FI}=12 & \rightarrow & 3\times \left\{4(1,2)\right\}
\ ,
\nonumber
\\
b)\ n_3^{FI}=6\ ,\,\ \ n_2^{FI}=18 & \rightarrow & 
3\times \left\{(3,1)+(\bar 3,1)+6(1,2)\right\}
\ ,
\nonumber
\\
c)\ n_3^{FI}=12\ ,\ n_2^{FI}=24 & \rightarrow &
 3\times \left\{2[(3,1)+(\bar 3,1)]+8(1,2)\right\}
\ ,
\nonumber 
\\
d)\ n_3^{FI}=18\ ,\ n_2^{FI}=30 & \rightarrow & 
3\times \left\{3[(3,1)+(\bar 3,1)]+10(1,2)\right\}
\label{extra5}
\ .
\eea
Thus
for a given Fayet-Iliopoulos scale, $M_{FI}$, 
each one of the four patterns in eq. (\ref{extra5})
will give rise to a different value
for $g_{GUT}$. Adjusting $M_{FI}$ appropriately,
we can always get $M_{GUT}\approx g_{GUT}\times 5.27\cdot 10^{17}$ GeV.
In particular this is so for 
$M_{FI}\approx 2\times 10^{16}$ GeV as shown in Fig.~2. 
It is remarkable that this number is within the allowed
range 
for the Fayet--Iliopoulos breaking scale as discussed above.
For the pattern in 
Fig.~2 corresponding to case a) we have 
$g_{GUT}\approx 1.1$, 
and therefore $M_{GUT}\approx 5.8\cdot 10^{17}$ GeV.


Of course, we cannot claim to have obtained the 
Heterotic String unification scale until we have shown
that the coupling $\alpha_1$ joins the other two couplings
at $M_{GUT}$.
The analysis becomes more involved now 
and a detailed account of this issue 
can be found in ref.~\cite{prediction}.
Let us just mention that the fact that the
normalization constant, $C$, of the $U(1)_Y$ hypercharge generator
is not fixed in these constructions 
as in the case of grand unified theories (e.g. for
$SU(5)$, $C^2=3/5$) is crucial in order to obtain the unification
with the other couplings.

\section{Phenomenology of this scenario}

The main characteristic of the scenario studied in the previous Section,
is the presence at low energy of extra matter. In particular, we have
obtained that three generations of Higgses and vector-like colour
triplets are necessary.

Since more Higgs particles than in the MSSM are present,
there will be of course a much richer phenomenology. 
Note for instance that the presence of six Higgs doublets
implies the existence of sixteen physical Higgs bosons,
eleven of them are neutral and five charged.  
On the other hand, it is well known that dangerous
flavour-changing neutral currents (FCNCs) may appear when
fermions of a given charge receive their mass through couplings
with several Higgs doublets.
This situation might be present here since we have three generations
of supersymmetric Higgses.
One approach in order to solve this potential problem is
to assume that the extra Higgses are
sufficiently  massive.
In this case
the actual lower bound 
on Higgs masses depends on the particular texture
choosen for the Yukawa matrices, but can be as 
low as 120--200 GeV.

Concerning the three generations of vector-like colour triplets, say 
$D$ and $\overline D$, 
they should acquire masses above the experimental limit ${O}$(200
GeV).
This is possible, in principle,
through 
couplings with some of the extra singlets with vanishing
hypercharge, 
say $N_i$, 
which are usually 
left at low energies, even after the Fayet--Iliopoulos breaking.
Thus couplings $N_iD\overline D$ might be present.
From the electroweak symmetry breaking, the fields $N_i$
a VEV might
develop. 

Before concluding, a few comments about the hypercharges of the
extra colour triplets are necessary. 
For the models studied in refs.~\cite{Font,Giedt2} they have
non-standard fractional electric charge,
$\pm 1/15$ and $\pm 1/6$ respectively.
In fact, the existence of this kind of matter is a generic property
of the massless spectrum of supersymmetric models.
This means that they have necessarily 
colour-neutral fractionally charged states, since
the triplets bind with the ordinary quarks.
For example, the model with triplets with electric charge
$\pm 1/6$ will have mesons and baryons with charges 
$\pm 1/2$ and $\pm 3/2$.
On the other hand,
the model studied in ref.~\cite{Casas2}
has `standard' extra triplets, i.e. with electric charges
$\mp 1/3$ and $\pm 2/3$; these will therefore give rise
to colour-neutral integrally charged states.
For example, a $d$-like quark $D$ forms
states of the type $u\overline D$, $uuD$, etc.

A detailed discussion about the stability of these charged states,
how to solve possible conflicts with cosmological bounds, and
their production modes
can be found in ref.~\cite{prediction}.

\section{Final comments}

We have attacked the problem of the unification of gauge couplings in
Heterotic String constructions. 
We have obtained that 
$\alpha_3$ and $\alpha_2$ cross at the right scale
when a certain type of extra matter is present. In this sense 
three families of supersymmetric Higgses and
vector-like colour triplets might be observed in forthcoming experiments.
The unification with 
$\alpha_1$ is obtained if the model has the
appropriate normalization factor of the hypercharge.

Let us recall that
although we have been working with explicit orbifold examples, our arguments
are quite general and can be used for other 
schemes where
the Standard Model gauge group with three generations of particles
is obtained,
since extra matter and
anomalous $U(1)$'s are generically present in compactifications of the
Heterotic String.


\begin{thebibliography}{0}





\bibitem{prediction} C. Mu\~noz, `A kind of prediction from
superstring model building',
{\it JHEP} {\bf 12} (2001) 015.








































\bibitem{Casas2} J.A. Casas and C. Mu\~noz,
`Three generation $SU(3)\times SU(2)\times U(1)_Y$ models from orbifolds',
{\it Phys. Lett.} {\bf B214} (1988) 63.

\bibitem{Font}
A. Font, L.E. Ib\'a\~nez, H.P. Nilles and F. Quevedo, 
`Yukawa couplings in degenerate orbifolds: towards a realistic
$SU(3)\times SU(2)\times U(1)$ superstring',
{\it Phys. Lett.} {\bf B210} (1988) 101.


\bibitem{Katehou}
J.A. Casas, E.K. Katehou and C. Mu\~noz, 
`$U(1)$ charges in orbifolds: anomaly cancellation and phenomenological
consequences',
{\it Nucl. Phys.} {\bf B317} (1989) 171.


%
\bibitem{Giedt2} J. Giedt, `Spectra in standard-like $Z_3$
orbifold models', hep-th/0108244.
%


\bibitem{Kaplu} V.S. Kaplunovsky,
`One loop threshold effects in string unification',
{\it Nucl. Phys.} {\bf B307} (1988) 145, 
{\it Erratum, ibid.} {\bf B382} (1992) 436.


\bibitem{rges} J.A. Casas and C. Mu\~noz,
`Restrictions on realistic superstring models from 
renormalization group equations',
{\it Phys. Lett.} {\bf B214} (1988) 543.






























































































































































\end{thebibliography}
\end{document}